\documentclass[a4paper,11pt]{article}
\usepackage{fullpage}

\newtheorem{theorem}{Theorem}
\newtheorem{lemma}[theorem]{Lemma}

\begin{document}

\title{$r$-indexing Wheeler graphs}
\author{Travis Gagie}
\date{\today}
\maketitle

\begin{abstract}
Let $G$ be a Wheeler graph and $r$ be the number of runs in a Burrows-Wheeler Transform of $G$, and suppose $G$ can be decomposed into $\upsilon$ edge-disjoint directed paths whose internal vertices each have in- and out-degree exactly 1.  We show how to store $G$ in $O (r + \upsilon)$ space such that later, given a pattern $P$, in $O (|P| \log \log |G|)$ time we can count the vertices of $G$ reachable by directed paths labelled $P$, and then report those vertices in $O (\log \log |G|)$ time per vertex.
\end{abstract}

\section{Introduction}
\label{sec:introduction}

Gagie, Manzini and Sir\'en~\cite{gagie2017wheeler} defined a Wheeler graph to be a directed multigraph whose edges are labelled with characters from a totally-ordered alphabet and whose vertices can be totally ordered such that those with in-degree 0 precede those with positive in-degree and, for any pair of edges $e = (u, v)$ and $e' = (u', v')$ labelled $a$ and $a'$ respectively,
\begin{itemize}
\item if $a \prec a'$ then $v < v'$,
\item if $a = a'$ and $u < u'$ then $v \leq v'$.
\end{itemize}
We call such an order on the vertices a Wheeler order.

Let $G$ be a Wheeler graph with Wheeler order $\pi$.  A Burrows-Wheeler Transform (BWT) of $G$ according to $\pi$ is a permutation of $G$'s edge labels such that, for any pair of edges $e = (u, v)$ and $e' = (u', v')$ labelled $a$ and $a'$ respectively, if $u < u'$ then $a$ precedes $a'$ in that permutation.  For convenience, we assume that the labels of each vertex's out-edges appear in the order in $\pi$ of their destinations.  Notice there may be many BWTs for $G$ because it may have many Wheeler orders.

Let $B$ be a BWT of $G$ according to $\pi$.  By the definition of a Wheeler graph, for any pattern $P$ over the alphabet of edge labels, the vertices reachable by directed paths labelled $P$ form an interval in $\pi$.  Moreover, if we store a rank data structure for $B$ and partial sum data structures for the frequencies of the distinct edge labels and the vertices' in- and out-degrees, then given $P$ we can find its interval in $O (|P| \log \log |G|)$ time.  Let $r$ is the number of runs (i.e., maximal non-empty unary substrings) in $B$ and suppose $G$ can be decomposed into $\upsilon$ edge-disjoint directed paths whose internal vertices each have in- and out-degree exactly 1.  Then these data  structures take a total of $O (r + \upsilon)$ space, measured in words.

Let $D$ be a such decomposition of $G$ and $n$ be the number of vertices in $G$, and assume the vertices are assigned numeric identifiers from 0 to $n - 1$ such that if $(u, v)$ is an edge and neither $u$ nor $v$ is an endpoint of a path in $D$, and $u$ has identifier $i$, then $v$ has identifier $i + 1$.  Notice these identifiers are not necessarily the vertices' ranks in $\pi$.  For convenience, we assume that even though $G$ is a multigraph, the number of edges is polynomial in $n$, so $\log \log |G| = O (\log \log n)$.  In this paper we show how, still using $O (r + \upsilon)$ space, after we have found the interval for $P$ we can then report the vertices in it using $O (\log \log n)$ time for each one.

We first prove a generalization of Bannai, Gagie and I's version~\cite{bannai2020refining} of Policriti and Prezza's Toehold Lemma~\cite{policriti2018lz77}, that lets us report the last vertex in the interval for $P$.  We then define a generalization of K\"arkk\"ainen, Manzini and Puglisi's $\phi$ function~\cite{karkkainen2009permuted}, that maps each vertex's identifier to the identifier of its predecessor in $\pi$.  Finally, we give a generalization of a key lemma behind Gagie, Navarro and Prezza's $r$-index~\cite{gagie2020fully}, that lets us compute our generalized $\phi$ function with $O (r + \upsilon)$-space data structures.  Combined, these three results yield a generalized $r$-index for Wheeler graphs.

\section{Generalized Toehold Lemma}
\label{sec:toehold}

For any pattern $P [0..m - 1]$, the interval for the empty suffix $P [m..m - 1]$ of $P$ is all of $\pi$, because every vertex is reachable by an empty path.  Assume we have found the interval $\pi [s_{i + 1}, e_{i + 1}]$ for $P [i + 1..m - 1]$ and now we want to find the interval $\pi [s_i, e_i]$ for $P [i..m - 1]$.  With the partial sum data structure for the vertices' out-degrees, in $O (\log \log n)$ time we can find the interval in $B$ containing the labels of the edges leaving the vertices in $\pi [s_{i + 1}, e_{i + 1}]$.

By the definition of a Wheeler graph, the edges labelled with the first and last occurrences of $P [i]$ in that interval in $B$, lead to the first and last vertices in the interval $\pi [s_i, e_i]$ for $P [i..m - 1]$.  Using the partial sum data structures for the frequencies of the distinct edge labels and the vertices in-degrees, in $O (\log \log n)$ time we can find the ranks $s_i$ and $e_i$ in $\pi$ of those first and last vertices in $\pi [s_i, e_i]$.  It follows that in $O (\log \log n)$ time we can find $\pi [s_i, e_i]$ from $\pi [s_{i + 1}, e_{i + 1}]$; therefore, by induction, we can find the interval for $P$ in $O (|P| \log \log n)$ time.  We can count the vertices in that interval in the same asymptotic time by simply returning the size of the interval.

To be able to find the identifier of the last vertex in the interval for $P$, for each edge $(u, v)$ we store $u$'s and $v$'s identifiers if any of the following conditions hold:
\begin{itemize}
\item $(u, v)$'s label $a$ is the last label in a run in $B$;
\item either $u$ or $v$ is an endpoint of a path in $D$;
\item the vertex that follows $u$ in $\pi$ has out-degree 0.
\end{itemize}
We store a select data structure for $B$, a bitvector marking the labels $a$ in $B$ for whose edges $(u, v)$ we have $u$'s and $v$'s identifiers stored, and a hash table mapping the position in $B$ of each marked label $a$ to the identifiers of its edge's endpoints.  This again takes a total of $O (r + \upsilon)$ space.

By querying the rank data structure, the select data structure, the bitvector and the hash table in that order, we can find the identifier of the vertex reached by the edge labelled by the last copy of $P [m - 1]$ in $B$.  By the definition of a Wheeler graph, this is the last vertex in the interval $\pi [s_{m - 1}, e_{m - 1}]$ for $P [m - 1]$.  Assume we have found the interval $\pi [s_{i + 1}, e_{i + 1}]$ for $P [i + 1..m - 1]$ and the identifier of the last vertex $u$ in that interval, and now we want to find the interval $\pi [s_i, e_i]$ for $P [i..m - 1]$ and the identifier of the last vertex $v$ in that interval.  We can find $\pi [s_i, e_i]$ as described above, so we need only say how to find $v$'s identifier.

With the partial sum data structure on the vertices' out-degree and the rank data structure, in $O (\log \log n)$ time we can check whether $u$ has an outgoing edge labelled $P [i]$.  If it does then, of all its out-edges labelled $P [i]$, the one whose label appears last in $B$ goes to $v$.  By our assumption of how the vertices are assigned their identifiers, if neither $u$ nor $v$ are endpoints of a path in $D$, then $v$'s identifier is $u$'s identifier plus 1.  If either $u$ or $v$ is an endpoint of a path in $D$, then we have $v$'s identifier stored and we can use the hash table to find it from the position in $B$ of the last label $P [i]$ on one of $u$'s out-edges, again in $O (\log \log n)$ time.

If $u$ does not have an outgoing edge labelled $P [i]$ then we can use the rank data structure to find the last copy of $P [i]$ in $B$ that labels an edge leaving a vertex in $\pi [s_{i + 1}, e_{i + 1}]$.  By the definition of a Wheeler graph, this edge $(u', v)$ goes to $v$.  Unlike in a BWT of a string, however, its label may not be the end of a run in $B$: $u$ could have out-degree 0, $u'$ could immediately precede $u$ in $\pi$ and the last of its outgoing edges' labels in $B$ could be a copy of $P [i]$, and the first label in $B$ of an outgoing edge of the successor of $u$ in $\pi$ could also be a copy of $P [i]$.  This is why we store $v$'s identifier if the vertex that follows $u'$ in $\pi$ has out-degree 0.  If $(u', v)$'s label is the end of a run in $B$, of course, then we also have $v$'s identifier stored.  In both cases we use $O (\log \log n)$ time, so from the interval $\pi [s_{i + 1}, e_{i + 1}]$ for $P [i + 1..m - 1]$ and the identifier of the last vertex $u$ in that interval, in $O (\log \log n)$ time we can compute the interval $\pi [s_i, e_i]$ for $P [i..m - 1]$ and the identifier of the last vertex $v$ in that interval.  Therefore, by induction, in $O (|P| \log \log n)$ time we can find the interval for $P$ and the identifier of the last vertex in that interval.

\begin{lemma}
\label{lem:toehold}
We can store $G$ in $O (r + \upsilon)$ space such that in $O (|P| \log \log n)$ time we can find the interval for $P$ and identifier of the last vertex in that interval.
\end{lemma} 

\section{Generalized $\phi$}
\label{sec:phi}

For a string $S$, the function $\phi$ takes a position $i$ in $S$ and returns the starting position of the suffix of $S$ that immediately precedes $S [i..|S| - 1]$ in the lexicographic order of the suffixes.  In other words, $\phi$ takes the value in some cell of suffix array of $S$ and returns the value in the preceding cell.  Given a pattern $P$, if we can find the interval of the suffix array containing the starting positions of occurrences of $P$ in $S$, and the entry in the last cell in that interval, then by iteratively applying $\phi$ we can report the starting positions of all the occurrences of $P$.  This is the idea behind the $r$-index for strings, which uses a lemma saying it takes only space proportional to the number of runs in the BWT of $S$ to store data structures that let us evaluate $\phi$ in $O (\log \log |S|)$ time.

We generalize $\phi$ to Wheeler graphs by redefining it such that it takes the identifier of some vertex $u$ in $G$ and returns the identifier of the vertex that immediately precedes $u$ in $\pi$.  (For the purposes of this paper, it is not important how $\phi$ behaves when given the identifier of the first vertex in $\pi$.)  Given a pattern $P$, if we can find the interval in $\pi$ containing the vertices in $G$ reachable by directed paths labelled $P$, and the identifier of the last vertex in that interval, then by iteratively applying $\phi$ we can report the identifiers of all those vertices.

Let $J$ be the set that contains $u$'s identifier if and only if any of the following conditions hold:
\begin{itemize}
\item $u$ has out-degree not exactly 1;
\item $u$ has a single outgoing edge $(u, v)$ but $v$ has in-degree not exactly 1;
\item the predecessor $u'$ of $u$ in $\pi$ has out-degree not exactly 1;
\item $u'$ has a single outgoing edge $(u', v')$ but $v'$ has in-degree not exactly 1;
\item the edges $(u, v)$ and $(u', v')$ have different labels.
\end{itemize}
We store a successor data structure for $J$ and, if $u$'s identifier is in $J$, then we store with it as satellite data the identifier of $u$'s predecessor $u'$ in $\pi$.  Notice $u$'s identifier is in $J$ only if at least one of $u$ or $u'$ or $v$ or $v'$ is the endpoint of a path in $D$, or the label of $(u', v')$ is the the last in a run in $B$ and the label of $(u, v)$ is the first in the next run.  It follows that we can use $O (r + \upsilon)$ space for the successor data structure and have it support queries in $O (\log \log n)$ time.

Suppose we know the identifier of some vertex $u$ with identifier $i$ that is immediately preceded by $u'$ in $\pi$ with identifier $i'$.  If $u \in J$ then we have $i'$ stored as satellite data with $\succ (i) = i$.  If $u \not \in J$, then $u$ has a single outgoing edge $(u, v)$ and $u'$ has a single outgoing edge $(u', v')$ with the same label, say $a$, and $v$ and $v'$ each have in-degree exactly 1.  By our assumption on how the identifiers are assigned, the identifiers of $v$ and $v'$ are $i + 1$ and $i' + 1$ and, by the definition of a Wheeler graph, $v$ is immediately preceded by $v'$ in $\pi$.  It follows that if $i + \ell$ is the successor of $i$ then it has stored with it as satellite data $i' + \ell$, and so we can compute $\ell$ and then $i'$ in $O (\log \log n)$ time.

\begin{lemma}
\label{lem:phi}
We can store $G$ in $O (r + \upsilon)$ space such that we can evaluate $\phi$ in $O (\log \log n)$ time.
\end{lemma}

\section{Conclusion}
\label{sec:conclusion}

Combining Lemmas~\ref{lem:toehold} and~\ref{lem:phi}, we generalize the $r$-index from strings to Wheeler graphs:

\begin{theorem}
\label{thm:main}
We can store $G$ in $O (r + \upsilon)$ space such that later, given a pattern $P$,  in $O (|P| \log \log n)$ time we can count the vertices of $G$ reachable by directed paths labelled $P$, and then report those vertices in $O (\log \log n)$ time per vertex.
\end{theorem}

Since $\upsilon = 1$ for a single string labelling a simple path or cycle, Theorem~\ref{thm:main} gives the same $O (r)$ space bound and $O (|P| + k \log \log n)$ time bound we achieve with the $r$-index for strings, where $k$ is the number of occurrences.  Nishimoto and Tabei~\cite{nishimoto2020faster} recently improved the query time of the $r$-index for strings to $O (P + k \log \log n)$ --- or optimal $O (P + k)$ for polylogarithmic alphabets --- without changing the space bound, and we conjecture this is achievable also for $r$-indexes for Wheeler graphs.

\end{document}